\documentclass[a4paper]{aa}
\usepackage{epsfig}
\usepackage{psfig}

\def\degmark{^\circ}
\def \rsun {\ifmmode$R$_{\odot}\else R$_{\odot}$\fi}
\def \nh {N${\rm _H}$}
\def \hcm {\hbox {\ifmmode $ atoms cm$^{-2}\else atoms cm$^{-2}$\fi}}
\def \src {1E\,1048.1--5937}

\def\approxgt{\mathrel{\hbox{\rlap{\lower.55ex \hbox {$\sim$}}
        \kern-.3em \raise.4ex \hbox{$>$}}}}
\def\approxlt{\mathrel{\hbox{\rlap{\lower.55ex \hbox {$\sim$}}
        \kern-.3em \raise.4ex \hbox{$<$}}}}

\newcommand {\ginga} {{\it Ginga}}

\newcommand {\sax} {{\it BeppoSAX}}

\newcommand {\Msun} {{{\rm M$_{\odot}$}}}

\def\arcmin{\hbox{$^\prime$}}

\newcommand {\chisq} {$\chi ^{2}$}
\newcommand {\rchisq} {$\chi_{\nu} ^{2}$}

\newcommand{\mc}{\multicolumn}

\begin{document}

\thesaurus{ (08.14.1; 08.16.7; 13.25.5)}

\title{The two-component X-ray spectrum of the 6.4 s pulsar \src}

\author{T.~Oosterbroek\inst{1} \and A.N.~Parmar\inst{1}
\and S. Mereghetti\inst{2} \and G.L. Israel\inst{3}}

\institute
{Astrophysics Division, Space Science Department of ESA, 
ESTEC, P.O. Box 299, 2200 AG Noordwijk, The Netherlands
\and
IFCTR, via Bassini 15, I-20133 Milano, Italy
\and
Osservatorio Astronomico di Roma, via dell'Osservatorio 1, I-00040, 
Monteporzio Catone, Italy}
\offprints{T. Oosterbroek: toosterb@astro.estec.esa.nl}
\date{Received ; accepted}
\maketitle

\markboth{T. Oosterbroek et al.: A \sax\ observation of the X-ray pulsar 
\src}{T. Oosterbroek et al.: A \sax\ observation of the X-ray pulsar \src}

\begin{abstract}
  
  The 6.4~s X-ray pulsar \src\ 
  was observed by \sax\ in 1997 May.  
  This source belongs to the class of ``anomalous'' pulsars which have pulse
  periods in range 5--11~s, show no evidence of optical
  or radio counterparts, and exhibit long-term increases in pulse period. 
  The phase-averaged 0.5--10 keV spectrum can be described by an absorbed
  power-law and blackbody model. The best-fit
  photon index is 2.5$\pm$0.2 and the blackbody temperature and
  radius are 0.64$\pm$0.01~keV and $0.59 \pm 0.02$~km (for a distance
  of 3~kpc), respectively. 
  The detection of blackbody emission from this
  source strengthens the similarity with two of the more well 
  studied ``anomalous''
  pulsars, 1E\,2259+586 and 4U\,0142+614. There is no evidence for any
  phase dependent spectral changes.

  The pulse period of 
  $6.45026 \pm 0.00001$~s implies that \src\ 
  continues to spin-down, but at a slower rate than obtained from 
  the previous measurements in 1994 and 1996.


\end{abstract}

\keywords{stars: neutron -- pulsars: individual (\src) --
X-rays: stars}

\section{Introduction}
\label{sec:introduction}

\src\ is an unusual pulsar, with a   period of 6.4~s and a soft
spectrum, discovered during {\it Einstein}\/ observations of 
the Carina nebula (Seward et al.\ 1983).
The source has been spinning down for at least the last 17 years
(Mereghetti 1995;  Corbet \& Mihara 1997). 
The spectrum has been modeled by an absorbed
power-law with a photon index, $\alpha$, of $\sim$2--3 (Seward et al.\ 1983;
Corbet \& Mihara 1997). This spectral shape
is softer than those of typical high-luminosity X-ray pulsars.
Despite a small error box, no optical   
counterpart has been identified, with a limiting magnitude of $m_{V}\sim20$
which excludes the presence of a massive companion 
(Mereghetti et al.\ 1992). 
A recent observation with the {\it RossiXTE}\/ satellite provides
a strong upper limit to the projected X-ray semi-major axis
of 0.06~lt-s 
for orbital periods between 200~s and $\sim$1~day
(Mereghetti et al.\ 1997).
The lack of an optical counterpart 
and orbital Doppler shifts argue
against a binary model for \src, unless the companion has a
low mass. Mereghetti et al.\ (1997) find that a probable
upper limit to the mass of a Roche lobe filling 
main sequence companion is $\sim$0.3
\Msun. Masses up to $\sim$0.8 \Msun\ are allowed in the case of
a helium-burning companion filling its Roche lobe.

The properties described above are similar to those of a 
small number of
other X-ray pulsars with spin periods in the 5--11~s range 
(Mereghetti \& Stella 1995), 
such as  1E\,2259$+$586 and 4U\,0142+614.
These form a class of so-called ``anomalous'' pulsars, 
with clearly different properties from the   
majority of systems.
Although accretion from a very low mass companion cannot be excluded,
the lack of
evidence for a binary nature from any of these systems has stimulated models
where the X-ray emission originates from a compact object that is
not in an interacting binary system.
While an isolated, massive, white dwarf powered by the loss of rotational 
energy, as originally proposed for 1E\,2259$+$586 (Paczynski 1990; Usov 1994),
has been ruled out by the detection
of a large increase in the spin-down rate of \src\  (Mereghetti 1995),
other single object models, such as loss
of magnetic energy of a strongly magnetized neutron star (Thompson \&
Duncan 1993), or an isolated neutron star accreting from a circumstellar disk
(Corbet et al. 1995; van Paradijs et al. 1995),   
may be applicable.



We present a study of \src\,
based on data obtained with the \sax\ satellite.
We focus on the X-ray spectrum at energies $<$10 keV and on the pulse period
history. As with some of the other ``anomalous'' pulsars, we
find evidence for the presence of a blackbody spectral component. 
Since this component is not observed from 
the majority of other accreting X-ray pulsars, this
strengthens the similarity between
\src\ and the other better studied ``anomalous'' pulsars.
 

\section{Observations}
\label{sec:observations}

Results were obtained with the Low-Energy Concentrator Spectrometer 
(LECS; 0.1--10~keV; Parmar et al. 1997a) and Medium-Energy Concentrator
Spectrometer (MECS; 1.3--10~keV; Boella et al. 1997) instruments.
The MECS consists of three identical grazing incidence
telescopes with imaging gas scintillation proportional counters in
their focal planes.  The LECS uses an identical telescope as
the MECS, but utilizes an ultra-thin (1.25~$\mu$m) entrance window and
a driftless configuration to extend the low-energy response to
0.1~keV. The fields of view of the LECS and MECS are circular
with diameters of 37\arcmin\ and 56\arcmin, respectively.  In the
overlapping energy range, the position resolution of both instruments
is similar and corresponds to 90\% encircled energy within a radius of
2\farcm5 at 1.5~keV. At lower energies, the encircled energy is
proportional to ${\rm E^{-0.5}}$. In the central 10$'$ the 
LECS 0.1--10~keV and the MECS 1--11~keV background
counting rates are $9.7 \times 10^{-5}$~arcmin$^{-2}$~s$^{-1}$ and
$10.9 \times 10^{-5}$~arcmin$^{-2}$~s$^{-1}$ (for 2 MECS units), respectively.
 
\src\ was observed by \sax\
on 1997 May 10 01$^{\rm h}$ 24$^{\rm m}$ to May 11 15$^{\rm h}$
24$^{\rm m}$ UTC.
Good data were selected from time intervals when the minimum elevation angle
above the Earth's limb was $>$4$\degmark$ and when the instrument
configurations were nominal using the SAXDAS 1.2.0 data analysis package.  
This gives exposures of 80~ks for the MECS and 32~ks
for the LECS, which was only operated during satellite night-time.
One of the three MECS units failed on 1997 May 9 (one day before our
observation), and data were only obtained from the remaining two
MECS units.

\begin{table}
\caption[]{Spectral fit results for \src. 
Uncertainties are given at 68\% confidence for one parameter of interest.
The normalization is at 1~keV}
\begin{flushleft}
\begin{tabular}{ll}
\hline\noalign{\smallskip}
Parameter & Value \\
\noalign {\smallskip}
\hline\noalign {\smallskip}
power-law model         \\
$\alpha$              & $3.36 \pm 0.05$\\
\nh\ ($10^{22}$~\hcm) & $1.54 \pm 0.10$ \\
\rchisq (dof)               & 1.520 (304) \\
\noalign{\smallskip}
\hline
\noalign{\smallskip}
power-law \& blackbody model         \\
$\alpha$              & $2.52 \pm0.20$\\
kT${\rm _{bb}}$ (keV) & $0.64 \pm0.01$ \\
R${\rm _{bb}}^a$ (km) & $0.59\pm0.02$\\
\nh\ ($10^{22}$~\hcm) & $0.45 \pm 0.10$ \\
\rchisq (dof)               & 0.997 (302) \\
\hline
\noalign{\smallskip}
Broken power-law model         \\
$\alpha_{1}$              & $1.71\pm0.2$\\
$\alpha_{2}$              & $3.58\pm0.08$\\
Break Energy (keV)        & $1.71\pm0.2$ \\
\nh\ ($10^{22}$~\hcm)     & $0.47 \pm 0.07$ \\
\rchisq (dof)               & 1.015 (302) \\
\hline
\noalign{\smallskip}
Two blackbody model         \\
kT${\rm _{bb}}1$ (keV) & $0.58\pm0.02$ \\
R${\rm _{bb}}1^a$ (km) & $0.77\pm0.08$\\
kT${\rm _{bb}}2$ (keV) & $1.25 \pm0.12$ \\
R${\rm _{bb}}2^a$ (km) & $0.10\pm0.03$\\
\nh\ ($10^{22}$~\hcm) & $0.14 \pm 0.04$ \\
\rchisq (dof)               & 0.964 (302) \\
\hline
\noalign{\smallskip}
Cut-off power-law model         \\
$\alpha$               & $0.42\pm0.07$ \\
Cut-off energy (keV)   & $1.43\pm0.13$ \\
\nh\ ($10^{22}$~\hcm) & $0.42 \pm 0.07$ \\
\rchisq (dof)               & 1.041 (303) \\
\hline
\noalign{\smallskip}
\multicolumn{2}{l}{\footnotesize $^a$For a distance of 3~kpc}
\end{tabular}
\end{flushleft}
\label{tab:pulsar_fits}
\end{table}

\section{Results}
\subsection {Spectral analysis}
\label{subsec:src_spectrum}

Spectra were obtained centered on the pulsar position
using extraction radii of 8\arcmin\ and 4\arcmin\ for the LECS and
MECS, respectively. The appropriate LECS response matrix was generated
using LEMAT 3.4.0.
Background subtraction was performed
using standard blank field exposures.
The background subtracted \src\ count rates  
are 0.08 and 0.12~s$^{-1}$ in the LECS and MECS,
respectively.  Examination of the LECS spectrum reveals that the
pulsar is only detected above 0.5~keV, and data below this energy
are excluded. Similarly, the MECS fits are restricted to the
energy range 1.65--10~keV.
All the  spectra were rebinned to have $>$20 counts in each bin to allow
the use of $\chi^2$ statistics. 

\begin{figure*}
\centerline{\mbox{\psfig{figure=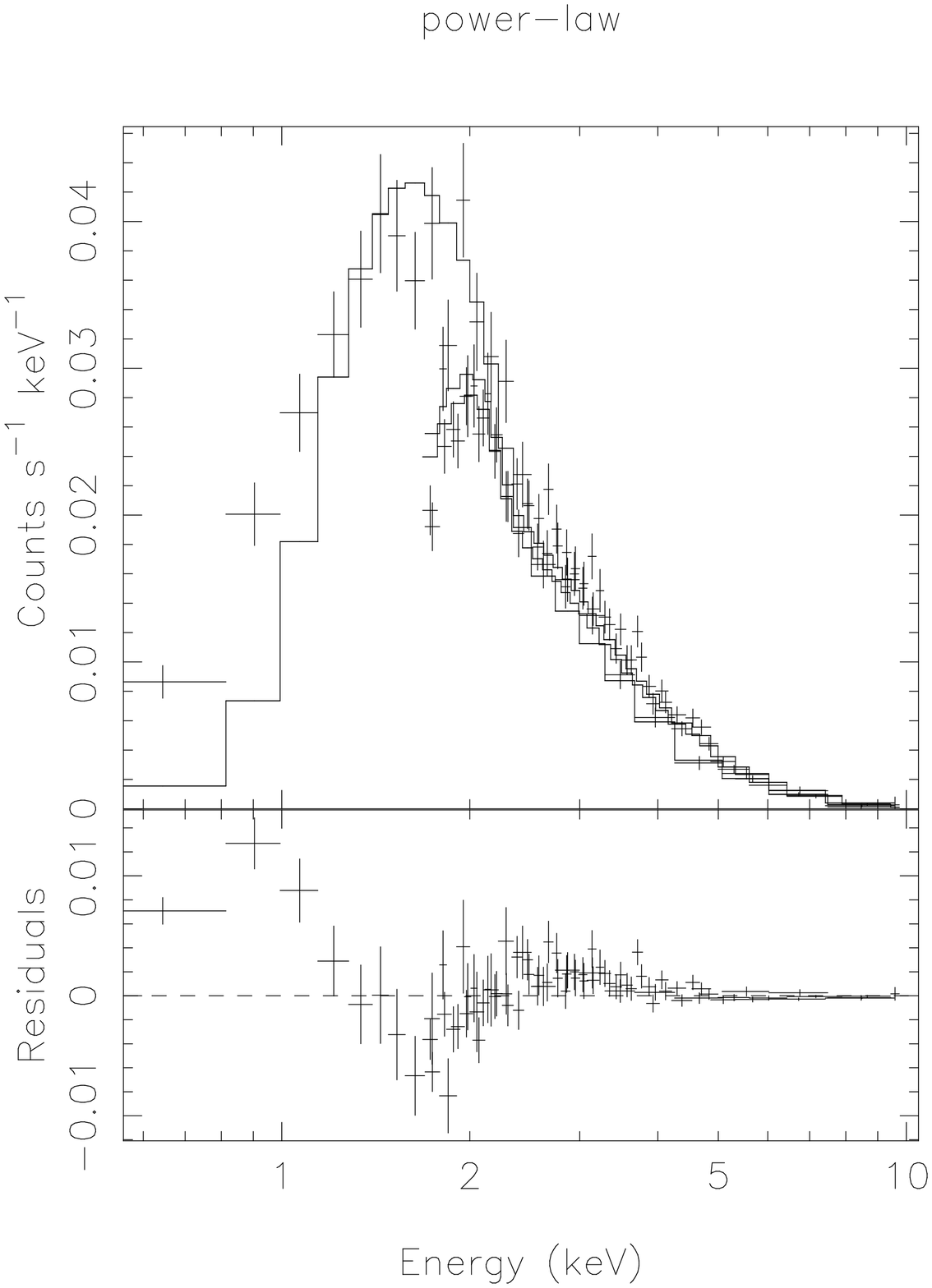,height=10cm}
\psfig{figure=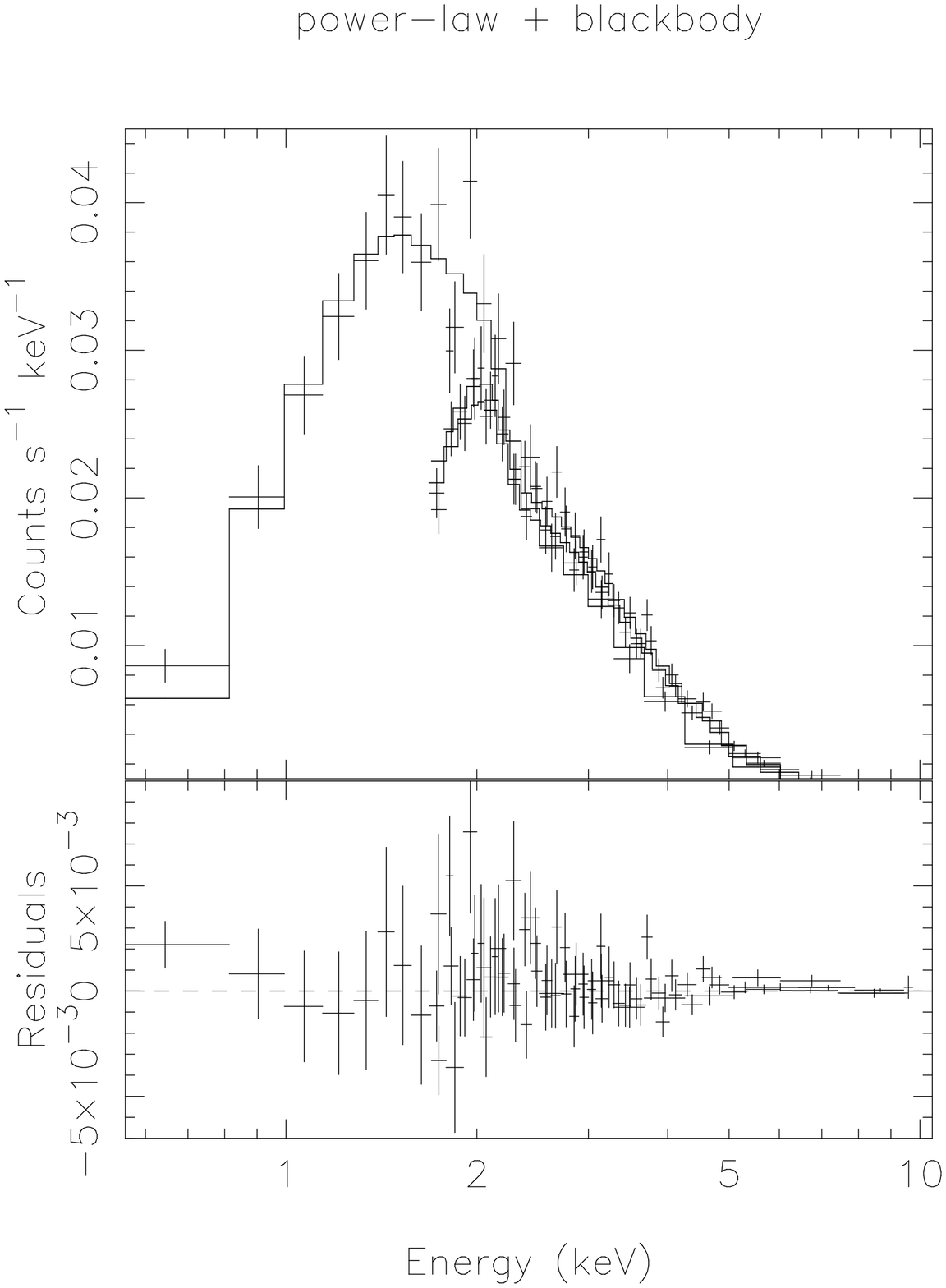,height=10cm}}}
\caption[]{The spectrum of \src\ obtained with the LECS and
two MECS units. The left panels show the spectrum and
the absorbed power-law model fit. 
Note the large residuals below 1.5 keV. The right panels show the
spectrum with the absorbed power-law
plus blackbody model fit. The units of the residuals are counts
s$^{-1}$ keV$^{-1}$}
\label{fig:spectrum}
\end{figure*}

In order to compare our results with previous observations, 
the spectra were first fit with two models: 
(i) an absorbed
power-law, and (ii) an absorbed power-law plus blackbody.
Although a single power-law with $\alpha$=4.04 satisfactorily
describes the MECS spectrum,
it is clearly inadequate when the LECS spectrum
is included, giving a \rchisq\ of 1.520 for 304 degrees of freedom (dof).
Figure \ref{fig:spectrum} shows the fit residuals before 
and after the  inclusion of the blackbody component.
The power-law plus blackbody model 
gives a \rchisq\ of 0.997 for 302 dof for a blackbody temperature, 
KT${\rm _{bb}}$,
of $0.64 \pm 0.01$~keV and radius, R${\rm _{bb}}$, of $0.59\pm0.02$~km for
a distance of 3~kpc.
The best-fit value of $\alpha$ is $2.52 \pm 0.20$ and low-energy
absorption, ${\rm N_H}$, equivalent to 
$(4.5 \pm 1.0)\times 10^{21}$~atoms~cm$^{-2}$ is required. 
An F-test  shows that the inclusion of the blackbody is highly significant  
(the reduction in \chisq\ has a formal probability
of $\sim$10$^{-28}$, corresponding to $\sim$11$\sigma$).
The best fit two-component spectrum is plotted 
in Fig.~\ref{fig:spectrum} and the corresponding parameters are given 
in Table \ref{tab:pulsar_fits}. 
The unabsorbed 2--10~keV source flux 
is 7.0$\times$10$^{-12}$ ergs cm$^{-2}$ s$^{-1}$,
with the blackbody component accounting for 55\% of the flux in this
energy range. For comparison, the results of fitting the same spectral
model to some other ``anomalous'' pulsars are given in Table~\ref{tab:bbprops}.
In order to investigate whether other simple spectral models can also
give acceptable representations of the \src\ spectrum, 
fits using three additional models
were also performed (see Table~\ref{tab:pulsar_fits}). All these models
gave fits of comparable quality as the power-law and blackbody combination.
This means that we are unable to meaningfully distinguish between the
different spectral models listed in Table~\ref{tab:pulsar_fits}, with
the exception of the power-law. However, in order to compare
our results with previous ones, we will use power-law and blackbody
fits in the subsequent analysis.

\subsection {Pulse timing and phase resolved spectroscopy}
\label{subsec:src_pulsetiming}

\begin{figure}
\centerline{\psfig{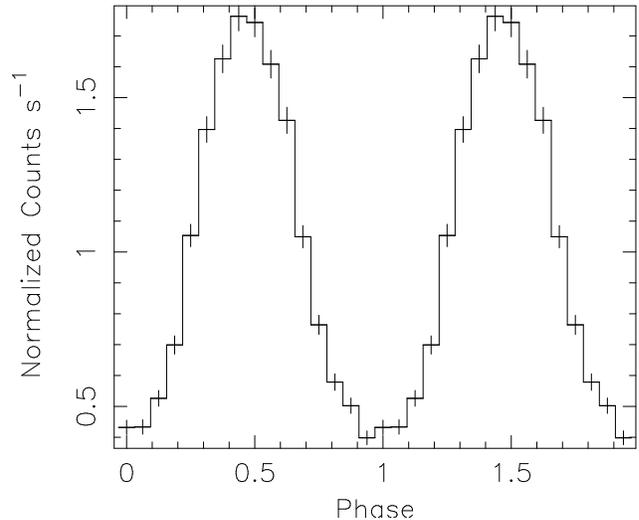}}
\caption[]{The 0.5--10~keV pulse profile of \src. For clarity,
the pulse profile is repeated}
\label{fig:profile}
\end{figure}

\begin{figure}
\centerline{\psfig{figure=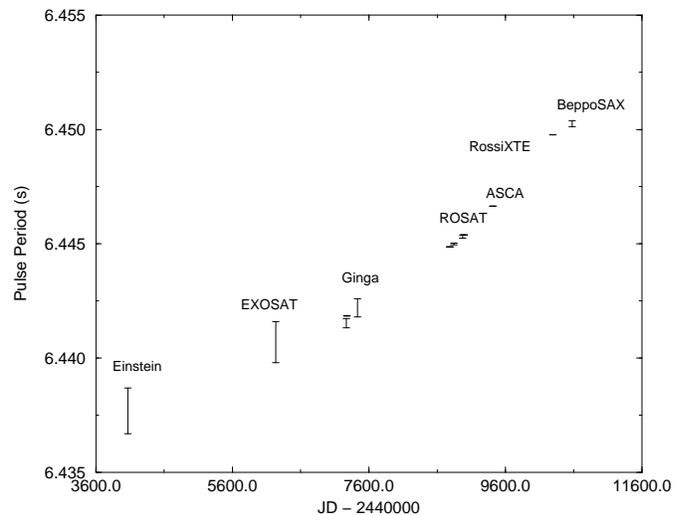,width=6.8cm,angle=-90,bbllx=93pt,bblly=41pt,bburx=564pt,bbury=648pt}}
\caption[]{The pulse period of \src\ as a function of time. Data have
been obtained from Mereghetti (1995), Corbet \& Mihara (1997) and
Mereghetti et al.\ (1997)}
\label{fig:history}
\end{figure}

The MECS counts were used to determine the \src\ 
pulse period, after correction of their 
arrival times to the solar system barycenter. 
The data were divided
into 10 time intervals (each with $\sim$950 counts) and for 
each interval the relative phase of the
pulsations determined. 
The phases of the 10 time intervals
were then fit with a linear function giving a best-fit period
of $6.45026 \pm 0.00001$~s. 
This value is plotted together with the 
previous measurements in Fig.\ \ref{fig:history}.
The 0.5--10~keV folded light curve of \src\ shows an approximately sinusoidal 
variation 
with an amplitude of $\sim$70\% (Fig.\ \ref{fig:profile}).
The pulsed fraction
as a function of energy was determined in the following way: 
The data for each energy range was folded at the
best-fit period and fit with a model consisting of a constant, $C$, 
and a sine term ($f(\phi) = C + A \sin(\phi + \phi_{0})$). 
The only free parameters in this fit are $C$
and the amplitude of the sine, $A$, since $\phi_{0}$ was determined 
from the total energy range data and was fixed during the
fits (there is no evidence for a change of $\phi_{0}$
with energy). 
We then determined the
pulsed fraction defined as the amplitude of the sine term ($A$)
divided by 
$C$ (a correction for background was applied). The results are given in
Table \ref{tab:fraction}. The correction for background is only 
important in the highest energy range. From this table we conclude
that the pulsed fraction is approximately constant over the whole
energy range.
We caution that the
systematic errors are largest in the highest energy bin (with the background
contributing 33\% of the mean count rate).

\begin{table}
\caption[]{The pulsed fraction as a function of energy as determined
from the MECS data}
\begin{tabular}{cc}
\hline
\mc{1}{c}{Energy Range} & \mc{1}{c}{Pulsed Fraction}\\
\mc{1}{c}{ (keV) } & \mc{1}{c}{(background corrected)}\\
\hline
1.0--1.5 & 0.76$\pm$0.08\\
1.5--2.0 & 0.74$\pm$0.05\\
2.0--2.5 & 0.77$\pm$0.04\\
2.5--4.0 & 0.84$\pm$0.03\\
4.0--6.0 & 0.79$\pm$0.05\\
6.0--10.0& 0.54$\pm$0.12\\
\hline
\end{tabular}
\label{tab:fraction}
\end{table}

\begin{table*}
\caption[]{The period history for \src}
\begin{tabular}{cr@{$\pm$}llcll}
\hline
\mc{1}{c}{Time}  & \mc{2}{c}{Period} & \mc{1}{c}{$\dot{P}^{a}$} & Pulsed
Fraction &\mc{1}{c}{Observatory} & Reference\\
\mc{1}{c}{(JD-2440000)} &\mc{2}{c}{(s)} & \mc{1}{c}{10$^{-4}$ s
yr$^{-1}$} &  & & \\
\hline
4068.7  & 6.4377 & 0.001      & \dots      & 0.68$\pm$0.07 & {\it Einstein} &
Seward et al.\ (1986)\\ 
6236.6  & 6.4407 & 0.0009     & 5$\pm$2 & 0.68          & EXOSAT &
Seward et al.\ (1986) \\
7267.0  & 6.44153 & 0.0002    & \dots     &               & {\it Ginga} &
Corbet \& Day (1990)\\
7277.0  & 6.44185 & 0.00001   & 4$\pm$3 & \dots$^{b}$     & {\it Ginga} & ''\\
7433.0  & 6.4422 & 0.0004     & \dots     &               & {\it Ginga} & ''\\
8787.24 & 6.444868 & 0.000007 & 7.29$\pm$0.03 & $\sim$0.70 & ROSAT &
Mereghetti (1995)\\
8844.48 & 6.44499 & 0.000034  & \dots          &            & ROSAT & ''\\
8972.74 & 6.44532 & 0.000072  & \dots         &            & ROSAT & ''\\
8994.80 & 6.445391 & 0.000013 & \dots          &            & ROSAT & ''\\
9417.0  & 6.446645 & 0.000001 & 10.30$\pm$0.04& 0.70$\pm$0.05$^{c}$ & ASCA &
Corbet \& Mihara (1997) \\
10294.5 & 6.449769 & 0.000004 & 12.99$\pm$0.02& \dots$^{d}$       &
{\it RossiXTE} & Mereghetti et al.\ (1997) \\
10579.31& 6.45026 & 0.000013  & 6.3$\pm$0.2   &
0.76$\pm$0.02  & {\it BeppoSAX} & This {\it paper} (see also Table \ref{tab:fraction})\\
\hline
\mc{7}{l}{$^{a}$~The values of $\dot{P}$ are calculated
by comparing the best pulse period determined by each satellite with
respect}\\
\mc{7}{l}{to that of the previous one, i.e.\ a long term average.}\\
\mc{7}{l}{$^{b}$~No value for the
pulsed fraction is reported for the Ginga observation due to
source confusion.}\\
\mc{7}{l}{$^{c}$~Obtained from Fig.\ 2 in
Corbet \& Mihara (1997)}\\
\mc{7}{l}{$^{d}$~Not reported.}\\
\end{tabular}
\label{tab:pdata}
\end{table*}

\begin{figure}
\centerline{\psfig{figure=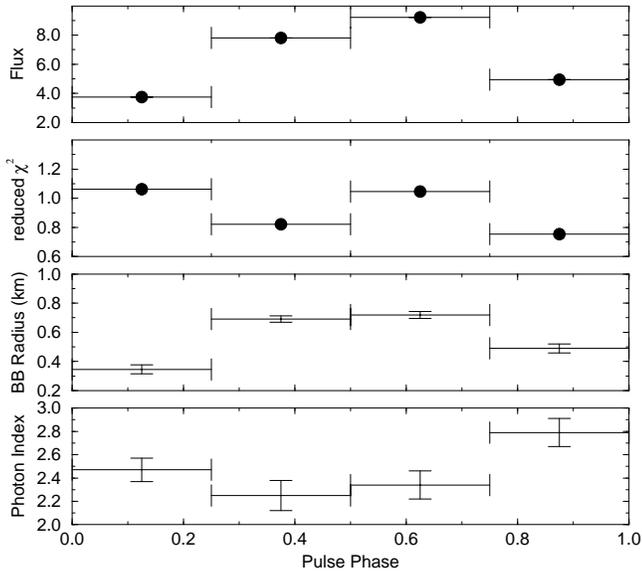,width=8.5cm,bbllx=12pt,bblly=21pt,bburx=563,bbury=511}}
\caption[]{Spectral fit  parameters as a function of pulse phase.
Errors are at 68\% confidence for 1 parameter of interest.
The blackbody radius is for an assumed distance of 3 kpc, the
flux in the top-panel is in units of $10^{-12}$ ergs s$^{-1}$
cm$^{-2}$. Pulse phase is the same as in Fig.\ \ref{fig:profile}}
\label{fig:variation}
\end{figure}

A set of 4 phase-resolved spectra of the pulsar were
accumulated.
These 
were then fit with the power-law plus blackbody model  
used in Sect.~\ref{subsec:src_spectrum}, with \nh\ fixed at the 
phase-averaged best-fit value.
There are insufficient counts to simultaneously constrain both the 
power-law and blackbody components.
Initially, the blackbody spectral parameters were fixed at their 
phase-averaged best-fit values and only the power-law parameters were
allowed to vary. The resulting fits are unacceptable, primarily because the 
spectra obtained at the valleys are not well fit (\rchisq=2.2 for
79 dof) due to the fixed blackbody component contributing
too much flux.
The fits were then repeated with the blackbody normalization allowed
to vary, giving acceptable values of \rchisq\ (second panel of
Fig.\ \ref{fig:variation}).

The best-fit values of $\alpha$ obtained with a variable blackbody
normalization 
are shown in Fig.\ \ref{fig:variation} (bottom panel) and reveal a marginal
phase dependence (\chisq\ of 11.1 for 3 dof,
corresponding to the 1\% level). A phase dependence of the blackbody
radius is evident.
The variation in $\alpha$ 
may be caused by inadequacies in the model,
or correlations between the
derived fit-parameters. 
The phase-resolved spectra were also fit 
with both $\alpha$ and
kT${\rm _{bb}}$ fixed and both normalizations free. This also gives
acceptable values of \rchisq\, which are almost identical to those
obtained from the fit described above.
The derived blackbody radii are not significantly different from those
obtained with the variable power-law index fits
(see Fig.\ \ref{fig:variation}).
We conclude that there is no evidence
for any phase dependent changes in either $\alpha$ or kT${\rm _{bb}}$.

\begin{table*}
\caption[]{The parameters of the ``anomalous'' pulsars}
\begin{center}
\begin{tabular}{llllllll}
\hline
 & P   & kT$_{\rm bb}$  & R$_{\rm bb}$ & \underline{L$_{\rm bb}$} & $\alpha$ &
f$_{\rm pulse}^{a}$ & References\\
Source & (s) & (keV) & (km) & L$_{\rm tot}$ & & \\       
\hline
4U\,0142+61 & 8.69 & 0.39 & 2.4d$_{\rm 1 kpc}$ & 0.4 & 3.67$\pm$0.09 &
$\sim$10\% & White et al.\ (1996)\\
RX J0720.4-3125$^{b}$ & 8.38 & 0.079$\pm$0.004$^{c}$  & \dots & \dots & 
\dots & \dots & Haberl et al. (1996) \\
1E\,1048-5937     &  6.45  & 0.64$\pm$0.01  & 0.59$\pm$0.02 d$_{\rm 3 kpc}$ & 0.55  & 2.5$\pm$0.2  & $\sim$70\% \\
1RXS J170849.0-400910 & 11.00 & 0.41$\pm$0.03 & 4$\pm$0.6 d$_{\rm 10 kpc}$ & 0.17 & 2.9$\pm$0.3 &
$\sim$30\% & Sugizaki et al.\ (1997)\\
1E\, 1841-045 & 11.77 & \dots & \dots & \dots & 3.0$\pm$0.2 & $\sim$35\% & Vasisht \& Gotthelf (1997)\\
1E\,2259+586 & 6.98 & 0.44$\pm$0.01 &  3.3$\pm$0.3 d$_{\rm 4 kpc}$  &   0.4
& 3.93$\pm$0.09 & $\sim$30\% & Corbet et al.\ (1995);\\
  & & & & & & & Parmar et al.\ (1997b)\\
\hline
\multicolumn{7}{l}{$^{a}$ Amplitude of pulsed emission}\\
\mc{7}{l}{$^{b}$ Membership of ``anomalous'' pulsar class uncertain}\\
\mc{7}{l}{$^{c}$ Obtained from one component blackbody fit}\\
\end{tabular}
\label{tab:bbprops}
\end{center}
\end{table*}

\begin{figure}
\centerline{\psfig{figure=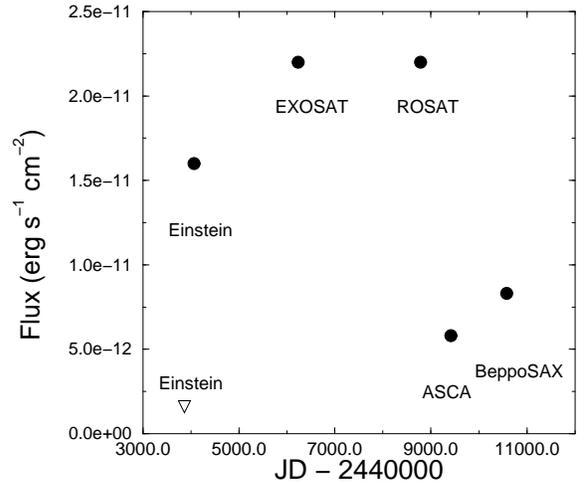,width=9cm}}
\caption[]{The 2--10 keV flux of \src\ as a function of
time. Data are from Seward et al.\ (1986), Mereghetti
(1995) and Corbet \& Mihara (1997). The ASCA and
\sax\ values are for a power-law fit (to be consistent with
the other measurements); 
the values obtained with 2-component fits are $\sim$15\% lower.
The first point (open triangle)
is the upper limit reported in Seward et al.\ (1986), where a short
discussion about the uncertainties in this value can be found.
ROSAT and {\it Einstein} fluxes in the 2--10 keV band are taken from
Mereghetti (1995) and Seward et al.\ (1986)}
\label{fig:fluxhistory}
\end{figure}

\section {Discussion}
\label{subsec:discussion}

The 0.5--10~keV spectrum of \src\ can be described by 
the sum of an absorbed power-law and blackbody models.  
The existence of a blackbody component in \src\ was first
suggested by the ASCA results of
Corbet \& Mihara (1997). 
However, these authors are unable to clearly discriminate between
this two component model, which gives a \rchisq\ of 0.94 for 332 dof, 
and an absorbed power-law which also provides
an acceptable description of the ASCA data with a \rchisq\ of
1.02 for 337 dof.
The \sax\ results reported here clearly require that the \src\ spectrum
differs significantly from an absorbed power-law. This deviation
in consistent with a blackbody, but we cannot exclude the possibility that
it has another form. A discussion about the physical interpretation of
the alternate models can be found in White et al.\ (1996). This
discussion is applicable to \src,
since its spectral shape is roughly similar to that of 4U\,0142+614.
The reason for the significant \sax\ detection of spectral complexity
is probably related to the combination of good energy resolution and
extended low energy coverage of the LECS, together with 
the longer \sax\ exposure. 

1E\,2259+586  (Corbet et al.\ 1995) and 4U\,0142+614 (White et al.\ 1996)
have also been successfully fit with absorbed power-law and blackbody
spectral models.
Since the majority of X-ray pulsar spectra are fit by absorbed power-law
models in the 0.5--10~keV energy range (e.g., White et al. 1983),
our results strengthen the similarity between \src\ 
and the other ``anomalous'' pulsars.
Table \ref{tab:bbprops} is a compilation of the spectral properties 
of these ``anomalous'' pulsars.
The blackbody component in these systems has been interpreted as 
evidence for quasi-spherical accretion onto an isolated neutron 
star formed after common envelope evolution and
spiral-in of a massive X-ray binary (White et al.\ 1996).  
In this case, the accretion flow results from the remaining part of
the massive star's envelope and may consist of two components
(Ghosh et al. 1997).  
A low-angular momentum component gives rise to the
blackbody emission from a considerable fraction of the neutron star surface,
while a high-angular momentum component forms an accretion disk and is
responsible for the power-law emission and the long term
spin-down evolution. 

The area for the blackbody emitting surface obtained for \src\
($\sim$0.59 d$_{\rm 3kpc}$ km) is smaller than 
those of the other ``anomalous'' systems.
However, the distance to \src\ is poorly constrained and
the assumed distance of 3~kpc could be considered as a 
lower limit 
since the  measured ${\rm N_H}$ implies that it lies behind the
Carina Nebula at 2.8~kpc (Seward et al.\ (1986).
White et al.\ (1996) propose that the low pulsed fraction
observed from 4U\,0142+614 results from the large polar
cap area in this system. This is consistent with the small polar
cap area and high pulsed fraction reported here for \src\, but not with
1E\,2259+586 and 1RXS{\thinspace}J170849.0$-$400910, which both have 
large radii and
a moderate pulsed fraction (see Table \ref{tab:bbprops}).
Interestingly, 
the best-fit blackbody temperature of 0.64~keV is somewhat 
higher than for 1E\,2259+586 and 4U\,0142+614 which 
may be related to the smaller area of the polar caps.
Summarizing, we find that for \src\ (i) the blackbody temperature is
higher, (ii) the blackbody radius is smaller, (iii) the power-law
index is smaller (i.e. the spectrum is harder), and (iv) the pulsed
fraction is higher than for the other ``anomalous'' pulsars (with the
exception of RX{\thinspace}J0720.4$-$3125 where the blackbody parameters 
are obtained from a one
component fit).

The phase-resolved spectra
are not well fit with a constant contribution from the blackbody
component. This is unsurprising given the large pulse amplitude
($\sim$70\%), together with the large blackbody contribution
(55\%; 2--10~keV) to the phase averaged flux. 
The probable constancy of the pulsed fraction over the 1--10~keV
energy range and the lack of any spectral
dependence on pulse phase (see Sect.\ \ref{subsec:src_pulsetiming}) 
imply that the {\it whole} spectrum is
varying in a similar manner; i.e. the pulsed component cannot be
attributed solely to either the blackbody or the power-law components.
Furthermore, the long term variations in source flux 
(Fig.\ \ref{fig:fluxhistory})
and the approximate constancy of the pulsed fraction (Table~\ref{tab:pdata}) 
during this time again implies that either the luminosities of the
two components are correlated, or that the underlying spectral shape is a more
complex ``single component'' that happens to mimic a power-law and
a blackbody. 

The spin-down rate of \src\ obtained from ROSAT and ASCA observations 
(Mereghetti 1995; Corbet \& Mihara 1997), 
showed an 80\% increase with respect to the value of
$\sim5 \times 10^{-4}$ s yr$^{-1}$ measured before 1988 with
\ginga\ and EXOSAT.  This increasing trend was further extended 
by the {\it RossiXTE}\/ observation of 1996 July ($P$ = $6.449769 \pm
0.000004$ s, Mereghetti et al.\ 1997), 
yielding a $\dot{P}$ of 13$\times10^{-4}$ s yr$^{-1}$ after the ASCA 
measurement.
The pulse period obtained with \sax\ lies significantly below
the linear extrapolation from the previous two measurements, 
indicating, for the first time in \src, a decrease in the
overall spin-down rate.

Table \ref{tab:pdata} and Fig.\
\ref{fig:fluxhistory} indicate that 
there is no clear correlation between the
observed long term spin-down rate ($\dot{P}$) and the 2--10 keV 
source flux. At first
sight this argues against an accretion hypothesis, but since
$\dot{P}$ is a long-term average and the flux an instantaneous
measurement this comparison may not be valid. 
A much better comparison would be between the average flux and
$\dot{P}$ during the same time interval. Unfortunately, due to the
faintness of the source, no such comparison is available.
In view of the uncertainties in the average flux a good measure of the
time variability of the source on timescales of months to years would
be useful.

\begin{acknowledgements}
  The \sax\ satellite is a joint Italian--Dutch programme.
  T. Oosterbroek acknowledges an ESA Fellowship. We thank the staff
  of the \sax\ Science Data Center for their support. We thank R.\
  Corbet for discussions and the anonymous referee for helpful
  suggestions.
\end{acknowledgements}

\end{document}